\documentclass[a4paper]{article}
\usepackage{spconf}

\usepackage{amssymb, amsmath}
\usepackage{amsfonts}
\usepackage{bbm}
\usepackage{xspace}
\usepackage{bm}

\usepackage{cite}
\usepackage{epsfig}
\usepackage{cases}
\usepackage{psfrag}
\usepackage{enumerate}

\usepackage{makeidx}
\usepackage{supertabular}

\newtheorem{theorem}{Theorem}

\newenvironment{textbmatrix}{   \setlength{\arraycolsep}{2.5pt}%
                                                                \big[\begin{matrix}}{\end{matrix}\big]%
                                                                \raisebox{0.08ex}{\vphantom{M}}}

\def\be{\begin{equation}}
\def\ee{\end{equation}}
\def\een{\nonumber \end{equation}}
\def\mat{\begin{bmatrix}}
\def\emat{\end{bmatrix}}
\def\btm{\begin{textbmatrix}}
\def\etm{\end{textbmatrix}}

\def\ba#1\ea{\begin{align}#1\end{align}}
\def\bs#1\es{\begin{split}#1\end{split}}
\def\bg#1\eg{\begin{gather}#1\end{gather}}
\def\bi#1\ei{\begin{itemize}#1\end{itemize}}

\newcommand{\safemath}[2]{\newcommand{#1}{\ensuremath{#2}\xspace}}

\safemath{\interior}{\mathrm{Int}}                       % interior of a set
\safemath{\dfn}{:=}                                                     % definition
\safemath{\dirac}{\delta}                                       % Dirac delta

\safemath{\SNR}{\text{\sc snr}}                                 % signal to noise ratio
\safemath{\No}{N_0}                                                     % noise spectral density
\safemath{\Es}{E_s}                                                     % energy per symbol
\safemath{\Eb}{E_b}                                                     % energy per bit
\safemath{\EbNo}{\frac{\Eb}{\No}}
\safemath{\EsNo}{\frac{\Es}{\No}}

\DeclareMathOperator{\CHop}{\ensuremath{\mathbb{H}}} % channel operator
\safemath{\tvir}{h_{\CHop}}                                     % time-varying impulse response
\safemath{\tvtf}{L_{\CHop}}                                     %       transfer function
\safemath{\spf}{S_{\CHop}}
\safemath{\bff}{H_{\CHop}}                                      %       function

\safemath{\ircf}{R_{h}}                                         % impulse response correlation fn.
\safemath{\scf}{R_{S}}                                          % scattering function
\safemath{\tfcf}{R_{L}}                                         % time-frequency correlation fn.
\safemath{\bfcf}{R_{H}}                                         % bi-frequency correlation fn.
\safemath{\mi}{I}                                                       % muttal information
\safemath{\capacity}{C}                                         % capacity

\newcommand{\pdf[2]}{f_{{#1}}\left({#2}\right)}                    % probability density function
\safemath{\uniform}{\mathcal{U}}                        % uniform distribution
\safemath{\normal}{\mathcal{N}}                         % normal distribution
\safemath{\circnorm}{\mathcal{CN}}                      % circ. symm. normal
\safemath{\mchain}{\leftrightarrow}                     % Markov chain

\safemath{\dB}{\,\mathrm{dB}}
\safemath{\dBm}{\,\mathrm{dBm}}
\safemath{\Hz}{\,\mathrm{Hz}}
\safemath{\kHz}{\,\mathrm{kHz}}
\safemath{\MHz}{\,\mathrm{MHz}}
\safemath{\GHz}{\,\mathrm{GHz}}
\safemath{\s}{\,\mathrm{s}}
\safemath{\ms}{\,\mathrm{ms}}
\safemath{\mus}{\,\mathrm{\mu s}}
\safemath{\ns}{\,\mathrm{ns}}
\safemath{\meter}{\,\mathrm{m}}
\safemath{\km}{\,\mathrm{km}}
\safemath{\mm}{\,\mathrm{mm}}
\safemath{\cm}{\,\mathrm{cm}}
\safemath{\m}{\,\mathrm{m}}
\safemath{\W}{\,\mathrm{W}}
\safemath{\J}{\,\mathrm{J}}
\safemath{\K}{\,\mathrm{K}}
\safemath{\bit}{\,\mathrm{bit}}
\safemath{\nW}{\,\mathrm{nW}}
\safemath{\muW}{\,\mathrm{$\mu$W}}
\safemath{\Watt}{\,\mathrm{W}}

\safemath{\define}{\triangleq}                  % definition

\safemath{\equivalent}{\sim}
\safemath{\distas}{\sim}                                        % distributed according to

\safemath{\reals}{\mathbb{R}}
\safemath{\positivereals}{\mathbb{R}^{+}}
\safemath{\integers}{\mathbb{Z}}
\safemath{\posint}{\mathbb{Z}_{+}}
\safemath{\naturals}{\mathbb{N}}
\safemath{\complexset}{\mathbb{C}}

\safemath{\setA}{\mathcal{A}}
\safemath{\setB}{\mathcal{B}}
\safemath{\setC}{\mathcal{C}}
\safemath{\setD}{\mathcal{D}}
\safemath{\setE}{\mathcal{E}}
\safemath{\setF}{\mathcal{F}}
\safemath{\setG}{\mathcal{G}}
\safemath{\setH}{\mathcal{H}}
\safemath{\setI}{\mathcal{I}}
\safemath{\setJ}{\mathcal{J}}
\safemath{\setK}{\mathcal{K}}
\safemath{\setL}{\mathcal{L}}
\safemath{\setM}{\mathcal{M}}
\safemath{\setN}{\mathcal{N}}
\safemath{\setO}{\mathcal{O}}
\safemath{\setP}{\mathcal{P}}
\safemath{\setQ}{\mathcal{Q}}
\safemath{\setR}{\mathcal{R}}
\safemath{\setS}{\mathcal{S}}
\safemath{\setT}{\mathcal{T}}
\safemath{\setU}{\mathcal{U}}
\safemath{\setV}{\mathcal{V}}
\safemath{\setW}{\mathcal{W}}
\safemath{\setX}{\mathcal{X}}
\safemath{\setY}{\mathcal{Y}}
\safemath{\setZ}{\mathcal{Z}}
\safemath{\emptySet}{\varnothing}

\safemath{\bma}{\mathbf{a}}
\safemath{\bmb}{\mathbf{b}}
\safemath{\bmc}{\mathbf{c}}
\safemath{\bmd}{\mathbf{d}}
\safemath{\bme}{\mathbf{e}}
\safemath{\bmf}{\mathbf{f}}
\safemath{\bmg}{\mathbf{g}}
\safemath{\bmh}{\mathbf{h}}
\safemath{\bmi}{\mathbf{i}}
\safemath{\bmj}{\mathbf{j}}
\safemath{\bmk}{\mathbf{k}}
\safemath{\bml}{\mathbf{l}}
\safemath{\bmm}{\mathbf{m}}
\safemath{\bmn}{\mathbf{n}}
\safemath{\bmo}{\mathbf{o}}
\safemath{\bmp}{\mathbf{p}}
\safemath{\bmq}{\mathbf{q}}
\safemath{\bmr}{\mathbf{r}}
\safemath{\bms}{\mathbf{s}}
\safemath{\bmt}{\mathbf{t}}
\safemath{\bmu}{\mathbf{u}}
\safemath{\bmv}{\mathbf{v}}
\safemath{\bmw}{\mathbf{w}}
\safemath{\bmx}{\mathbf{x}}
\safemath{\bmy}{\mathbf{y}}
\safemath{\bmz}{\mathbf{z}}

\bmdefine{\biad}{a}
\bmdefine{\bibd}{b}
\bmdefine{\bicd}{c}
\bmdefine{\bidd}{d}
\bmdefine{\bied}{e}
\bmdefine{\bifd}{f}
\bmdefine{\bigd}{g}
\bmdefine{\bihd}{h}
\bmdefine{\biid}{i}
\bmdefine{\bijd}{j}
\bmdefine{\bikd}{k}
\bmdefine{\bild}{l}
\bmdefine{\bimd}{m}
\bmdefine{\bind}{n}
\bmdefine{\biod}{o}
\bmdefine{\bipd}{p}
\bmdefine{\biqd}{q}
\bmdefine{\bird}{r}
\bmdefine{\bisd}{s}
\bmdefine{\bitd}{t}
\bmdefine{\biud}{u}
\bmdefine{\bivd}{v}
\bmdefine{\biwd}{w}
\bmdefine{\bixd}{x}
\bmdefine{\biyd}{y}
\bmdefine{\bizd}{z}

\bmdefine{\bixid}{\xi}
\bmdefine{\bilambdad}{\lambda}
\bmdefine{\bimud}{\mu}
\bmdefine{\bithetad}{\theta}
\bmdefine{\biphid}{\phi}

\safemath{\bmia}{\biad}
\safemath{\bmib}{\bibd}
\safemath{\bmic}{\bicd}
\safemath{\bmid}{\bidd}
\safemath{\bmie}{\bied}
\safemath{\bmif}{\bifd}
\safemath{\bmig}{\bigd}
\safemath{\bmih}{\bihd}
\safemath{\bmii}{\biid}
\safemath{\bmij}{\bijd}
\safemath{\bmik}{\bikd}
\safemath{\bmil}{\bild}
\safemath{\bmim}{\bimd}
\safemath{\bmin}{\bind}
\safemath{\bmio}{\biod}
\safemath{\bmip}{\bipd}
\safemath{\bmiq}{\biqd}
\safemath{\bmir}{\bird}
\safemath{\bmis}{\bisd}
\safemath{\bmit}{\bitd}
\safemath{\bmiu}{\biud}
\safemath{\bmiv}{\bivd}
\safemath{\bmiw}{\biwd}
\safemath{\bmix}{\bixd}
\safemath{\bmiy}{\biyd}
\safemath{\bmiz}{\bizd}

\safemath{\bmxi}{\bixid}
\safemath{\bmlambda}{\bilambdad}
\safemath{\bmmu}{\bimud}
\safemath{\bmtheta}{\bithetad}
\safemath{\bmphi}{\biphid}

\safemath{\bA}{\mathbf{A}}
\safemath{\bB}{\mathbf{B}}
\safemath{\bC}{\mathbf{C}}
\safemath{\bD}{\mathbf{D}}
\safemath{\bE}{\mathbf{E}}
\safemath{\bF}{\mathbf{F}}
\safemath{\bG}{\mathbf{G}}
\safemath{\bH}{\mathbf{H}}
\safemath{\bI}{\mathbf{I}}
\safemath{\bJ}{\mathbf{J}}
\safemath{\bK}{\mathbf{K}}
\safemath{\bL}{\mathbf{L}}
\safemath{\bM}{\mathbf{M}}
\safemath{\bN}{\mathbf{N}}
\safemath{\bO}{\mathbf{O}}
\safemath{\bP}{\mathbf{P}}
\safemath{\bQ}{\mathbf{Q}}
\safemath{\bR}{\mathbf{R}}
\safemath{\bS}{\mathbf{S}}
\safemath{\bT}{\mathbf{T}}
\safemath{\bU}{\mathbf{U}}
\safemath{\bV}{\mathbf{V}}
\safemath{\bW}{\mathbf{W}}
\safemath{\bX}{\mathbf{X}}
\safemath{\bY}{\mathbf{Y}}
\safemath{\bZ}{\mathbf{Z}}

\safemath{\bZero}{\mathbf{0}}

\bmdefine{\biAd}{A}
\bmdefine{\biBd}{B}
\bmdefine{\biCd}{C}
\bmdefine{\biDd}{D}
\bmdefine{\biEd}{E}
\bmdefine{\biFd}{F}
\bmdefine{\biGd}{G}
\bmdefine{\biHd}{H}
\bmdefine{\biId}{I}
\bmdefine{\biJd}{J}
\bmdefine{\biKd}{K}
\bmdefine{\biLd}{L}
\bmdefine{\biMd}{M}
\bmdefine{\biOd}{N}
\bmdefine{\biPd}{O}
\bmdefine{\biQd}{P}
\bmdefine{\biRd}{R}
\bmdefine{\biSd}{S}
\bmdefine{\biTd}{T}
\bmdefine{\biUd}{U}
\bmdefine{\biVd}{V}
\bmdefine{\biWd}{W}
\bmdefine{\biXd}{X}
\bmdefine{\biYd}{Y}
\bmdefine{\biZd}{Z}

\bmdefine{\biDelta}{\Delta}
\bmdefine{\biLambda}{\Lambda}
\bmdefine{\biPhi}{\Phi}
\bmdefine{\biSigma}{\Sigma}
\bmdefine{\biOmega}{\Omega}
\bmdefine{\biTheta}{\Theta}

\safemath{\bimA}{\biAd}
\safemath{\bimB}{\biBd}
\safemath{\bimC}{\biCd}
\safemath{\bimD}{\biDd}
\safemath{\bimE}{\biEd}
\safemath{\bimF}{\biFd}
\safemath{\bimG}{\biGd}
\safemath{\bimH}{\biHd}
\safemath{\bimI}{\biId}
\safemath{\bimJ}{\biJd}
\safemath{\bimK}{\biKd}
\safemath{\bimL}{\biLd}
\safemath{\bimM}{\biMd}
\safemath{\bimN}{\biNd}
\safemath{\bimO}{\biOd}
\safemath{\bimP}{\biPd}
\safemath{\bimQ}{\biQd}
\safemath{\bimR}{\biRd}
\safemath{\bimS}{\biSd}
\safemath{\bimT}{\biTd}
\safemath{\bimU}{\biUd}
\safemath{\bimV}{\biVd}
\safemath{\bimW}{\biWd}
\safemath{\bimX}{\biXd}
\safemath{\bimY}{\biYd}
\safemath{\bimZ}{\biZd}

\safemath{\bDelta}{\bielta}
\safemath{\bLambda}{\biLambda}
\safemath{\bPhi}{\biPhi}
\safemath{\bSigma}{\biSigma}
\safemath{\bOmega}{\biOmega}
\safemath{\bTheta}{\biTheta}

\safemath{\veca}{\bma}
\safemath{\vecb}{\bmb}
\safemath{\vecc}{\bmc}
\safemath{\vecd}{\bmd}
\safemath{\vece}{\bme}
\safemath{\vecf}{\bmf}
\safemath{\vecg}{\bmg}
\safemath{\vech}{\bmh}
\safemath{\veci}{\bmi}
\safemath{\vecj}{\bmj}
\safemath{\veck}{\bmk}
\safemath{\vecl}{\bml}
\safemath{\vecm}{\bmm}
\safemath{\vecn}{\bmn}
\safemath{\veco}{\bmo}
\safemath{\vecp}{\bmp}
\safemath{\vecq}{\bmq}
\safemath{\vecr}{\bmr}
\safemath{\vecs}{\bms}
\safemath{\vect}{\bmt}
\safemath{\vecu}{\bmu}
\safemath{\vecv}{\bmv}
\safemath{\vecw}{\bmw}
\safemath{\vecx}{\bmx}
\safemath{\vecy}{\bmy}
\safemath{\vecz}{\bmz}
\safemath{\vecZero}{\bZero}

\safemath{\vecxi}{\bmxi}
\safemath{\veclambda}{\bmlambda}
\safemath{\vecmu}{\bmmu}
\safemath{\vectheta}{\bmtheta}
\safemath{\vecphi}{\bmphi}

\safemath{\matA}{\bA}
\safemath{\matB}{\bB}
\safemath{\matC}{\bC}
\safemath{\matD}{\bD}
\safemath{\matE}{\bE}
\safemath{\matF}{\bF}
\safemath{\matG}{\bG}
\safemath{\matH}{\bH}
\safemath{\matI}{\bI}
\safemath{\matJ}{\bJ}
\safemath{\matK}{\bK}
\safemath{\matL}{\bL}
\safemath{\matM}{\bM}
\safemath{\matN}{\bN}
\safemath{\matO}{\bO}
\safemath{\matP}{\bP}
\safemath{\matQ}{\bQ}
\safemath{\matR}{\bR}
\safemath{\matS}{\bS}
\safemath{\matT}{\bT}
\safemath{\matU}{\bU}
\safemath{\matV}{\bV}
\safemath{\matW}{\bW}
\safemath{\matX}{\bX}
\safemath{\matY}{\bY}
\safemath{\matZ}{\bZ}
\safemath{\matZero}{\bZero}

\safemath{\matDelta}{\bDelta}
\safemath{\matLambda}{\bLambda}
\safemath{\matPhi}{\bPhi}
\safemath{\matSigma}{\bSigma}
\safemath{\matOmega}{\bOmega}
\safemath{\matTheta}{\bTheta}

\safemath{\matIdentity}{\matI}

\usepackage{algorithm,algorithmic}
\usepackage[usenames, dvipsnames]{color}
\usepackage{subfigure}

\safemath{\crb}{\mathsf{CRB}}
\safemath{\mse}{\mathsf{MSE}}
\safemath{\var}{{\rm{VAR}}}
\safemath{\cov}{{\rm{COV}}}
\safemath{\nvar}{\mathsf{nvar}}

\safemath{\dif}{\,\mathrm{d}}

\newfont{\bb}{msbm10 scaled 1100}

\begin{document}

\title{Energy Consumption in multi-user MIMO systems: \\ Impact of user mobility}

\name{Luca Sanguinetti$^{\dagger \ddagger}$ \qquad Aris L. Moustakas$^{\star\ddagger}$ \qquad Emil Bj{\"o}rnson$^{\Diamond\ddagger}$  \qquad M{\'e}rouane Debbah$^{\ddagger}$ \thanks{L. Sanguinetti has received funding from the People Programme (Marie Curie Actions) FP7 PIEF-GA-2012-330731 Dense4Green. E. Bj\"ornson is funded by an International Postdoc Grant 2012-228 from The Swedish Research Council. A. L. Moustakas is the holder of the DIGITEO "ASAPGONE" Chair. This research has been also supported by the ERC Starting Grant 305123 MORE and by the FP7 Network of Excellence in Wireless COMmunications NEWCOM\# (Grant agreement no. 318306).}
}
\address{
$^{\dagger}$\small{Dipartimento di Ingegneria dell'Informazione, University of Pisa, Pisa, Italy}\\
$^{\star}$\small{Department of Physics,
National \& Capodistrian University of Athens, Athens, Greece}\\
$^{\Diamond}$\small{ACCESS Linnaeus Centre, Department of Signal Processing, KTH Royal Institute of Technology, Stockholm, Sweden}\\
$^\ddagger$\small{Alcatel-Lucent Chair, Ecole sup{\'e}rieure d'{\'e}lectricit{\'e} (Sup{\'e}lec), Gif-sur-Yvette, France}}

\ninept
\maketitle

\begin{abstract}
In this work, we consider the downlink of a single-cell multi-user multiple-input multiple-output system in which zero-forcing precoding is used at the base station (BS) to serve a certain number of user equipments (UEs). A fixed data rate is guaranteed at each UE. The UEs move around in the cell according to a Brownian motion, thus the path losses change over time and the energy consumption fluctuates accordingly. We aim at determining the distribution of the energy consumption. To this end, we analyze the asymptotic regime where the number of antennas at the BS and the number of UEs grow large with a given ratio. It turns out that the energy consumption is asymptotically a Gaussian random variable whose mean and variance are derived analytically. These results can, for example, be used to approximate the probability that a battery-powered BS runs out of energy within a certain time period.
\end{abstract}

\section{Introduction}

The data traffic in cellular networks has increased exponentially for a long time and is expected to continue this trend, at least for the next five years \cite{Cisco11}. Currently, one of the biggest challenges related to the traffic growth is the increasing energy consumption of the cellular infrastructure equipments \cite{Fehske11}. This means that the energy consumption must be taken into account from the very beginning when designing cellular networks of the future. This is particularly important when deploying BSs in new rural regions of the world, where the electrical grid is unreliable or even non-existing. Off-grid deployments rely on combinations of diesel generators, batteries, and local energy harvesting  (e.g., from solar panels) \cite{Fehske11}. Since the supply of energy is either limited or fluctuates with the harvesting, it is of paramount importance to operate the BS such that it will not run out of energy, also known as power outage.

Most of the existing works dealing with the development of energy-efficient transmission schemes rely on extensive Monte-Carlo simulations that do not provide valuable insights on the interplay between the different system parameters and the impact of the propagation environment. To partially bridging this gap, in \cite{Xiang13} the authors make use of stochastic geometry to model the energy consumption of a cellular network where each UE is connected to its closest BS equipped with a single antenna. The energy consumption is expressed as a function of the distance between BSs and UEs, while taking into account the interference from other BSs. In \cite{Decreusefond13}, the authors go a step further in this direction when considering a refined energy consumption model that includes the energy of broadcast messages, traffic activity, and user mobility. Instead of relying on simulation results, as was done in \cite{Xiang13}, closed-form formulas for different statistical properties of the energy consumption are derived. {From a network deployment perspective, \cite{Emil13} shows how the number of BS antennas, number of active UEs, and the data rates can be analytically optimized for high energy efficiency. This is achieved using a refined energy consumption model where the three optimization variables appears explicitly.}

In this work, we consider the downlink of a single-cell multi-user multiple-input multiple-output (MIMO) system. The BS is equipped with an array of $N$ antennas and serve $K$ UEs simultaneously by using zero-forcing (ZF) precoding. The UEs are assumed to move around in the cell according to a Brownian motion model. Based on this mobility pattern, we aim at determining the statistical distribution of the energy consumption required to guarantee a given data rate at all UEs. By considering the asymptotic regime where $K$ and $N$ grow large with a given ratio, we prove that energy consumption statistics converge in distribution to a Gaussian random variable whose mean and variance can be analytically derived using random matrix theory tools and standard central limit theoretic results. It turns out in the large limit the variance of the energy consumption is dominated by the fluctuations induced by user mobility. The analytical expressions are shown to closely match the numerical results for different settings. Finally, we exemplify how the new statistical characterization can be used to characterize the probability that a battery-powered BS runs out of energy within a certain time period.

The following notation is used throughout this work. $\mathbb{E}_{\mathbf{z}}[\cdot]$, $\cov_{\mathbf{z}}[\cdot]$ and $\var_{\mathbf{z}}[\cdot]$ indicate that the expectation, covariance and variance are respectively computed with respect to $\mathbf{z}$. The notation $||\cdot||$ stands for the Euclidean norm whereas $J_n(\cdot)$ denotes the $n$-order Bessel function. We call $\mathbf{I}_K$ the $K \times K$ identity matrix and $\delta(t)$ the Dirac delta function. We denote by $Q(z)$ the Gaussian tail function and use $Q^{-1}(z)$ to indicate its inverse.

\section{System and signal model}

%\subsection{System model}

We consider the downlink of a single-cell multi-user MIMO system in which the BS makes use of $N$ antennas to communicate with $K$ single-antenna UEs. The $K$ active UEs change over time and are randomly selected from a large set of UEs that are moving around within the coverage cell $\mathcal A$ of area $A$. We assume that the user density $\nu$ and the number of UEs $K$ can increase arbitrarily, while the area $A$ is maintained fixed. This amounts to saying that $K/\nu$ is constant and equal to $A$. To simplify the computations, we assume that the UEs are uniformly distributed in a circular cell with radius
$R$ such that $A= \pi R^2$ and adopt a Brownian motion (or random walk mobility) model with diffusion constant $D$ and constrained in the circular region $\mathcal A$ \cite{Camp02asurvey}. The location of UE $k$ at time $t$ is denoted by $\mathbf{x}_k(t) \in \mathbb{R}^2$. The BS is located in the centre of the cell and its $N$ transmit antennas are adequately spaced apart such that the channel components to any UE are uncorrelated. We assume that $N$ increases as $K$ becomes larger while the ratio $K/N$ is kept constant and equal to $c$ with $0<c<1$. Perfect channel state information is assumed to be available at the BS and the same rate is guaranteed to each UE \cite{Emil13}.

%\subsection{Signal model}
We call $\mathbf{s}(t) \in \mathbb{C}^{N\times 1}$ the signal transmitted at (slotted) time $t$ and denote by $\mathbf{G}(t) \in \mathbb{C}^{N\times N}$ its precoding matrix. We assume that $\mathbf{s}(t)$ {originates from a Gaussian codebook with zero mean and covariance matrix} $\mathbb{E}_{\mathbf{s}}[\mathbf{s}(t)\mathbf{s}^H(t)] = \mathbf{I}_K$. Letting $\mathbf{y}(t) \in \mathbb{C}^{K\times 1}$ be the vector collecting the samples received at the UEs, we may write
\begin{align}\label{y_k}
\mathbf{y}(t)= \mathbf{H}(t)\mathbf{G}(t)\mathbf{s}(t)+\mathbf{n}(t)
\end{align}
where $\mathbf{n}(t) \in \mathbb{C}^{K\times 1}$ is a circularly-symmetric complex Gaussian random vector with zero-mean and covariance matrix $\sigma^2\mathbf{I}_K$ and $\mathbf{H}(t)\in \mathbb{C}^{K\times N}$ is the channel matrix at time $t$. The ($k,n$)th entry $\left[\mathbf{H}(t)\right]_{k,n}$ accounts for the channel propagation coefficient between the $n$th antenna at the BS and the $k$th UE. In particular, we assume
\begin{align}\label{H_k,n}
\left[\mathbf{H}(t)\right]_{k,n}= \sqrt{g(\mathbf{x}_{k}(t))}\left[\mathbf{W}(t)\right]_{k,n}
\end{align}
where $g(\cdot)$ is the path-loss function and the entries $\left[\mathbf{W}(t)\right]_{k,n}$ account for the fast fading component and are modelled as independent and identically distributed circularly-symmetric complex Gaussian random variables with zero-mean and unit variances, i.e., $\left[\mathbf{W}(t)\right]_{k,n} \sim \mathcal {CN} (0,1)$. {The temporal correlations of $\mathbf{W}(t)$ are modelled according to the Jakes model} \cite{Jakes_book}. We assume that
\begin{align}\label{pathloss}
g(\mathbf{x}_{k}(t)) =\frac{1}{\left\|\mathbf{x}_{k}(t)\right\|^{\beta} + r_0^\beta }
\end{align}
with $\beta$ and $r_0$ being the path-loss exponent and some cutoff parameter, respectively. As seen, $g(\mathbf{x}_{k}(t))$ is assumed to be independent over $n$. This is a reasonable assumption since the distances between UEs and BS are much larger than the distance between the antennas.

{For analytical convenience, we consider the ZF precoding matrix} \begin{align}\label{G}
\mathbf{G}(t)=\sqrt{\rho} \,\mathbf{H}^H(t)\left(\mathbf{H}(t)\mathbf{H}^H(t)\right)^{-1}
\end{align}
where $\rho$ is a design parameter. Substituting $\mathbf{G}(t)$ into \eqref{y_k} the achievable data rate of the $k$th UE is
\begin{align}\label{rate}
r_k= \log_2\left(1+\frac{\rho}{\sigma^2}\right).
\end{align}
{Note that we assume that the same rate is achieved by each UE \cite{Emil13}. The extension to the case in which different rates are required by different UEs can be easily handled by combining ZF with a proper power allocation \cite{Bjornson2013d}.}

The power consumption $P(t) = \mathbb{E}_{\mathbf{s}}[||\mathbf{G}(t)\mathbf{s}(t)||^2]$ at time $t$ is given by
\begin{align}\label{P_T}
P(t)=\rho \,{\rm{tr}}\left(\left(\mathbf{H}(t)\mathbf{H}^H(t)\right)^{-1}\right)
\end{align}
while the energy consumption $E_T$ for a given time interval $[0,T]$ is
\begin{align}\label{E}
E_T=\int_0^T {P(t)dt} = \int_0^T \rho \,{\rm{tr}}\left(\left(\mathbf{H}(t)\mathbf{H}^H(t)\right)^{-1}\right)d t.
\end{align}

\section{Main results}

The energy $E_T$ is clearly a random quantity, which depends (through $\mathbf{H}(t)$) on the realizations of $\mathbf{W}(t)$ as well as on the user positions $\{\mathbf{x}_{k}(t)\}$ throughout the period $0 \leq t \leq T$. We aim at determining the statistics of $E_T$ in the large system limit, i.e., $K,N \rightarrow \infty$ with $K/N = c$. For notational convenience, we denote by $k_i$ the $i$th zero of the first derivative of $J_1(\cdot)$ and call
\begin{align}
\phi_i =2\int_0^1  J_0(k_{i}t) t^{\beta+1} dt.
\end{align}
Observe that the values of $\{k_i\}$ and $\{\phi_i\}$ can be calculated explicitly as illustrated in \cite{Gradshteyn_Ryzhik_book}.

The following theorem summarizes the main results of this work.

%which also can be calculated explicitly (see 6.561(1) in \cite{Gradshteyn_Ryzhik_book}).
\begin{theorem}
In the large system limit, if ZF precoding is used then the following convergence holds true:
\begin{align}\label{convergence}
\frac{E_T - \mathbb{E}\left[E_T\right] }{\sqrt{\var\left[E_T\right]}} \underset{K,N \to \infty}{\longrightarrow} \mathcal{N}(0,1)
\end{align}
where the mean is given by
\begin{align}\label{avg_pow}
\mathbb{E}\left[E_T\right] = T \frac{\rho c R^\beta}{1-c} \left(\frac{2}{\beta+2}+\frac{r_0^\beta}{R^\beta}\right)
\end{align}
while the variance depends on the user mobility model and takes the form
\begin{align}\label{var_pow}
\var\left[E_T\right] &= \frac{T R^{2}}{DK} \frac{\rho^2 c^2R^{2\beta}}{(1-c)^2}\Theta +O(K^{-2})
\end{align}
with $\Theta = \sum\nolimits_{i=1}^\infty \frac{2\phi^2_i}{ k^2_iJ_0^2(k_i)}\int_0^1{(1-e^{-\frac{k_i^2DTt}{R^2}})^2 dt}$.
%\begin{align}\label{var_pow.1}
%\Theta = \sum_{i=1}^\infty \frac{\phi^2_i}{D k^2_iJ_0^2(k_i)}\int_0^T\left(1-e^{-\frac{k_i^2Dt}{R^2}}\right)^2 dt.
%\end{align}
\end{theorem}

\subsection{Sketch of proof}
The complete proof of Theorem 1 is omitted for space limitations. In the sequel, we describe the main steps.

We begin by observing that when $K,N \rightarrow \infty$ with $0<c<1$, the average of $P(t)$ with respect to the fast fading channel $\mathbf{W}(t)$ hardens to a deterministic scalar given by\cite{Hachem2007_DeterministicEquivCertainFunctionalsRandomMatrices, Bai2004_CLT_covariance_matrices}
\begin{align}\label{avg_pow_conditioned}
\mathbb{E}_{{\bf W}}\left[P(t ) \right] - \frac{\rho c}{1-c} \frac{1}{K}\sum_{k=1}^K \frac{1}{g(\mathbf{x}_{k}(t))}\rightarrow  0.
%&= \frac{\rho}{1-c} \frac{1}{K}\sum_{k=1}^K \left(|{\bf x}_k(t)|^\beta+x_0^\beta\right).
\end{align}
%from which it follows that the average power is the sum of the powers needed by $K$ independent channels to invert the pathloss of each UE, with an additional penalty of $(1-c)^{-1}$ present due to the non-orthogonality between antenna channels.
Since the path-loss functions $g(\mathbf{x}_{k}(t))$ are independent of each other, from \eqref{avg_pow_conditioned} it follows that in the large system limit
\begin{align}\label{avg_pow_conditioned2}
\mathbb{E}_{{\bf x}, {\bf W}} \left[P(t) \right] - \frac{\rho c}{1-c}\mathbb{E}_{{\bf x}}\left[ \frac{1}{g(\mathbf{x}_{k}(t))}\right] \rightarrow  0
%&= \frac{\rho}{1-c} \left(\frac{2R^\beta}{\beta+2}+x_0^\beta\right)
\end{align}
from which using \eqref{pathloss} the result in \eqref{avg_pow} easily follows. Observe that the exceedingly simple form in (\ref{avg_pow}) hold true only for ZF precoding technique. It becomes more involved if other precoding techniques (such as the regularized ZF) are considered.

The variance of the energy $E_T$ can be rewritten (using the covariance decomposition formula) as ${\rm{VAR}}\left[E_T\right] =  A_1 + A_2$
%\begin{align}\label{var_E.1}
%{\rm{VAR}}\left[E_T\right] =  A_1 + A_2
%\end{align}
with
\begin{align}\label{A_1}
A_1 & = \int_{0}^T\int_{0}^T \mathbb{E}_{\mathbf{x}}\left[{\rm{COV}}_{\mathbf{W}}\left[ {P(t),P(t^\prime)} \right]\right]d td t^\prime \\ \label{A_2}
A_2 & = \int_{0}^T\int_{0}^T {\rm{COV}}_{\mathbf{x}}\left[\mathbb{E}_{\mathbf{W}}\left[{P(t)}\right],\mathbb{E}_{\mathbf{W}}\left[{P(t^\prime)}\right]\right]d td t^\prime.
\end{align}
%
%
%To quantify the fluctuations of $E_T$, we start observing that they are originated from two sources: the variability of the fading coefficients in ${\bf W}(t)$ and the user mobility ${\bf x}(t)$. In particular, the total variance can be written as a sum of the two variances, namely
%\begin{align}
%\var\left[E_T\right] = \var_{\bf W}\left[E_T\right]+ \var_{\bf x}\left[E_T\right]
%\end{align}
Let us start with the computation of $A_1$. For this purpose, we observe that according to the Jakes model \cite{Jakes_book}, the correlation time $\tau_d = 1/{f_d}$ (with $f_d$ being the Doppler frequency) is exceedingly small compared to the typical times related with the UE movements. This means that we can reasonably assume that the power is $\delta$-correlated or, equivalently, that $\cov_{{\bf W}}\left[P(t),P(t')\right]=\var_{{\bf W}}\left[P(t)\right] \tau_d \delta(t-t')$. As a result, the fluctuation term $A_1$ in the energy can be approximated as $A_1 \approx T\tau_d \mathbb{E}_{\mathbf{x}}\left[\var_{{\bf W}}\left[P(t)\right]\right]$.
%\begin{align}\label{eq:var_fading_energy}
%A_1 \sim T\tau_d \mathbb{E}_{\mathbf{x}}\left[\var_{{\bf W}}\left[P(t)\right]\right].
%\end{align}
From \cite{Maiwald2000_WishartMoments}, using \eqref{pathloss} we obtain
%\begin{align}\label{eq:var_fading_power}
%  \var_{{\bf W}}\left[P(t) \right] \mathop  \to \limits_{K,N \rightarrow \infty} \frac{c}{(1-c)^3}\frac{1}{K^3}\sum_{k=1}^K g_k^{-2}(t)
%\end{align}
%from which using \eqref{pathloss} yields
\begin{align}\label{eq:var_fading_power}
%\int_{0}^T\int_{0}^T  \mathbb{E}_{\mathbf{x}} & \left[{\rm{COV}}_{\mathbf{W}} \left[{P(t),P(t^\prime)}\right]\right]d td t^\prime  \mathop  \to \limits_{K,N \rightarrow \infty} \nonumber \\ & \frac{c}{(1-c)^3}\frac{1}{K^3}\sum_{k=1}^K \mathbb{E}_{\mathbf{x}}\left[g_k^{-2}(t)\right].
\mathbb{E}_{\mathbf{x}}\left[\var_{{\bf W}}\left[P(t)\right]\right] -\frac{\rho^2c^3}{(1-c)^3} \frac{T\tau_d}{K^2} \mathbb{E}_{{\bf x}}\left[ \frac{1}{g^2(\mathbf{x}_{k}(t))}\right] \rightarrow 0
\end{align}
with
\begin{align}
\mathbb{E}_{{\bf x}}\left[  \frac{1}{g^2(\mathbf{x}_{k}(t))}\right]=  \frac{R^{2\beta}}{2\beta+2} + \frac{4r_0^\beta R^\beta}{\beta+2} + r_0^{2\beta}
\end{align}
as obtained using \eqref{pathloss}. Plugging the above results into \eqref{A_1} reveals that $A_1$ decreases proportionally to $K^{-2}$ when $K \rightarrow \infty$.

We are now left with the computation of $A_2$ in \eqref{A_2}, which is due to the fluctuations induced by the UE mobility in the cell. In particular, it turns out that
\begin{align}\label{eq:var_brownian_power}
A_2  = \frac{\rho^2c^2}{K(1-c)^2} \int_{0}^T\int_{0}^T \cov_{\mathbf{x}}\left[\left\|\mathbf{x}(t)\right\|^\beta,\left\|\mathbf{x}(t^\prime)\right\|^\beta\right] dt dt'.
\end{align}
According to the Brownian motion model, averaging over the initial positions of the UE (or random walkers in the Brownian motion parlance) yields
\begin{align}\label{17}
&\cov_{\mathbf{x}}\left[\left\|\mathbf{x}(t)\right\|^\beta,\left\|\mathbf{x}(t^\prime)\right\|^\beta\right] = \\ \nonumber
&\iint_{{\mathbf x , \mathbf x^\prime} \in \mathcal A} \left\|\mathbf{x}\right\|^\beta\left\|\mathbf{x}^\prime\right\|^\beta \left(F(\mathbf{x},\mathbf{x}^\prime;t-t^\prime) - F(\mathbf{x},\mathbf{x}^\prime;t+t^\prime)\right) d{\bf x}d{\bf x}'
\end{align}
with $F({\bf x},{\bf x'};t)$ being the probability that a random walker at position ${\bf x'}$ reaches position ${\bf x}$ at time $t$. Denote now by $\xi$ the time of each step of the random walker (or, equivalently, the ``forgetting time''---the time after which the walker forgets his original direction) and call $\ell$ the spatial length of each step. For values of $t$ much larger than $\xi$, $F({\bf x},{\bf x'};t)$ can be obtained in the continuum limit by solving a diffusion equation, whose corresponding diffusion constant turns out to be equal to $D=\ell ^2/(4\xi)$. The diffusion equation can be solved very simply in the circular domain by providing the proper initial condition $F({\bf x},{\bf x}',t=0)=\delta({\bf x}-{\bf x}')$. To impose the condition that the random walker does not exit the domain, we need to set the derivative along the radial direction at the boundary equal to zero, i.e., ${\hat {\mathbf r}}^T\nabla{\left.F({\bf x},{\bf x}')\right|_{\|{\bf x}\|=R}}=0$ with ${\hat{\mathbf r}}$ being the unit vector along the radial direction.
%\textcolor{red}{One has to also set boundary conditions, which are chosen to be $\partial_rF({\bf x},{\bf x}')|_{\left\|{\bf x}\right\|=R}=0$. This amounts to saying that the radial derivative of $F({\bf x},{\bf x}')$ at the boundary vanishes, to ensure that the walker remains in the domain of the cell.}
As a result, an eigenfunction expansion for $F({\bf x},{\bf x}')$ can be obtained and used to compute \eqref{17} from which using \eqref{eq:var_brownian_power} the result in \eqref{var_pow} follows.

In writing  \eqref{var_pow} we have taken into account that $A_1$ is $\propto \tau_d T R^{2\beta}/K^{2} $ while $A_2$ is $\propto T R^{2\beta+2}/(DK)$. This difference in scaling arises from two different effects. Firstly, the extra factor of $K^{-1}$ in the former is due to the fact that the singular values of ${\bf W}(t)$ are strongly correlated, while the positions of the UEs are independent. Secondly, the decorrelation time for the fading is $\tau_d$, while that for the user mobility is $R^2/D$, which is much larger than $\tau_d$. Hence, both factors make the variability of the fast fading channel less important for the analysis.

%In writing this result, we have taken into account that the fast fading term in \eqref{eq:var_fading_power} is smaller than \eqref{eq:var_brownian_power} by a factor of $K^{-1}$. This is due to the fact that the singular values of the matrix ${\bf H}$ are strongly correlated \textcolor{red}{(due to level repulsion)}, while the positions of the UEs are independent. \textcolor{red}{There is an additional factor of $R^2/D\tau_D$, which accounts for the fact that the correlation time in the Brownian motion is $\sim R^2/D$, much larger compared to $\tau_D$ in the fast fading.} Both factors make the variability of the fast fading less important for this analysis.

From the results of Theorem 1, it is seen that both the mean and the variance of $E_T$ (at least for long enough time interval $T$ such that $DT\gg R^2$) are proportional to $T$. This means that the variability of energy consumption will be less important as $T$ becomes larger.

Finally, it can be shown that all higher-order moments of $E_T$ asymptotically vanish for large values of $K$.
%This happens taking into account either the fluctuations due to the fast fading or those induced by user mobility.
In particular, tools from random matrix theory \cite{Bai2004_CLT_covariance_matrices, Hachem2007_DeterministicEquivCertainFunctionalsRandomMatrices} can be used to control the variations of the fast fading channel while standard central limit theoretic results are needed to prove the convergence of the fluctuations arising from user movements.

\begin{figure}
  \begin{center}
    \psfrag{xlabel}[c][b]{{$\frac{E_T - \mathbb{E}\left[E_T\right] }{\sqrt{\var\left[E_T\right]}}$}}
    \psfrag{ylabel}[c][t]{{Frequency}}
    \psfrag{data1}[l][m]{\!\!\!\!\footnotesize{Histogram}}
    \psfrag{data2}[l][m]{\!\!\!\!\footnotesize{$\mathcal N(0,1)$}}
        \psfrag{title}[c][m]{\!\!\!\!\footnotesize{$T=10, K= 16, N=32, \beta=4$}}
    \includegraphics[width=0.85\columnwidth]{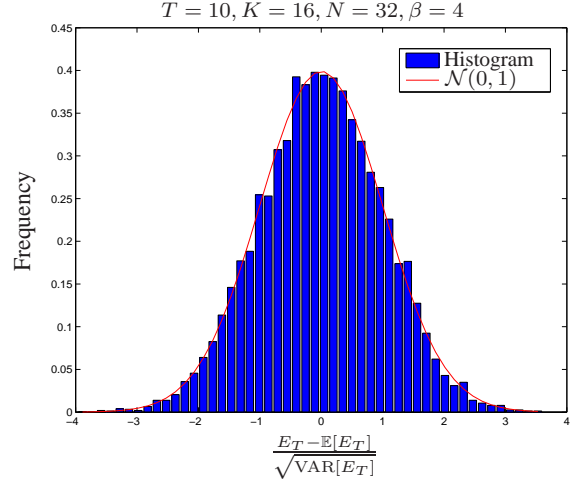}
  \caption{Histogram of $({E_T - \mathbb{E}\left[E_T\right] })/{\sqrt{\var\left[E_T\right]}}$ in comparison with the normal distribution $\mathcal N(0,1)$ for $T=10, K= 16, N=32$ and $\beta=4$.}\vskip-0.7cm
  \label{fig5}
  \end{center}
\end{figure}

\subsection{Numerical Validation}

The accuracy of the results in Theorem 1 are now validated numerically by Monte-Carlo simulations for 10000 different initial positions of the UEs. For illustration purposes, we simply set $R =1$, $r_0 = 0.1$, $\rho =1$, $\ell =0.05$ and $\xi = \ell^2 = 0.0025$.

Fig.~\ref{fig5} depicts the histogram of $({E_T - \mathbb{E}\left[E_T\right] })/{\sqrt{\var\left[E_T\right]}}$ when $T=10$, $K=16, N=32$ and $\beta =4$. Comparisons are made with the Gaussian distribution $\mathcal N(0,1)$. The good match between the two curves validates the results of Theorem 1.

\begin{figure}
  \begin{center}
    \psfrag{xlabel}[c][b]{{$\alpha$}}
    \psfrag{ylabel}[c][t]{{$\text{Pr}\left(\frac{E_T}{T} > \alpha \right)$}}
    \psfrag{x5}[c][m]{\footnotesize{$N=64$}}
    \psfrag{x6}[c][m]{\footnotesize{$N=32$}}
    \psfrag{data1}[l][m]{\!\!\!\!\footnotesize{$T=2$ - Sim.}}
    \psfrag{data2}[l][m]{\!\!\!\!\footnotesize{$T=2$ - Theory}}
    \psfrag{data3}[l][m]{\!\!\!\!\footnotesize{$T=10$ - Sim.}}
    \psfrag{data4}[l][m]{\!\!\!\!\footnotesize{$T=10$ - Theory}}
        \psfrag{title}[c][m]{\!\!\!\!\footnotesize{$K= 16, \beta=4$}}
    \includegraphics[width=0.95\columnwidth]{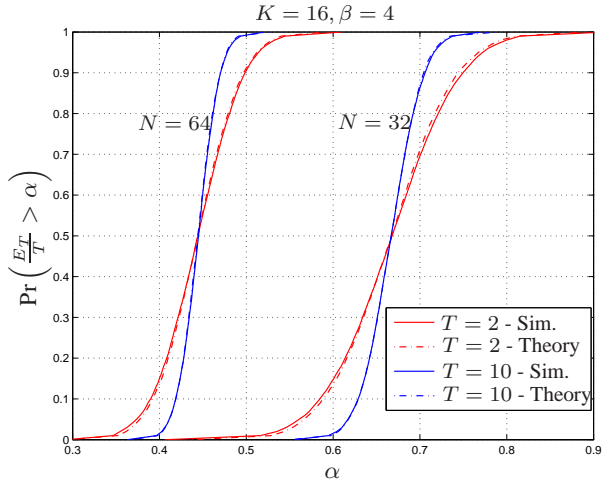}
  \caption{Outage probability $\text{Pr}\left(\frac{E_T}{T} > \alpha \right)$ for $T=2$ and $T=10$ with $K= 16, \beta=4$ and $N=32$ or $64$.}\vskip-0.6cm
  \label{fig1}
  \end{center}
\end{figure}

\begin{figure}
  \begin{center}
    \psfrag{xlabel}[c][b]{{$\alpha$}}
    \psfrag{ylabel}[c][t]{{$\text{Pr}\left(\frac{E_T}{T} > \alpha \right)$}}
        \psfrag{x5}[c][m]{\footnotesize{$\beta = 4$}}
    \psfrag{x6}[c][m]{\footnotesize{$\beta = 6$}}
    \psfrag{data1}[l][m]{\!\!\!\!\footnotesize{$T=2$ - Sim.}}
    \psfrag{data2}[l][m]{\!\!\!\!\footnotesize{$T=2$ - Theory}}
    \psfrag{data3}[l][m]{\!\!\!\!\footnotesize{$T=10$ - Sim.}}
    \psfrag{data4}[l][m]{\!\!\!\!\footnotesize{$T=10$ - Theory}}
        \psfrag{title}[c][m]{\!\!\!\!\footnotesize{$K= 64, N=128$}}
    \includegraphics[width=0.95\columnwidth]{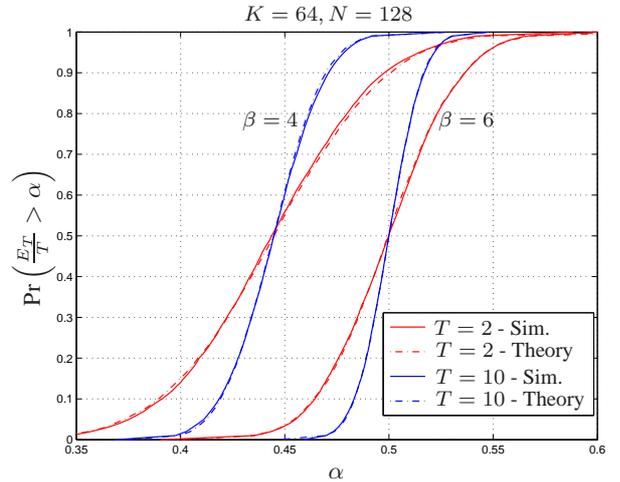}
  \caption{Outage probability $\text{Pr}\left(\frac{E_T}{T} > \alpha \right)$ for $T=2$ and $T=10$ with $K= 64, N=128$ and $\beta=4$ or $6$.}\vskip-0.6cm
  \label{fig3}
  \end{center}
\end{figure}

Figs.~\ref{fig1} and \ref{fig3} show the cumulative distribution function (CDF) of ${E_T}/{T}$ for different values of $K$, $N$, $T$ and $\beta$. The values of $\mathbb{E}\left[E_T\right]$ and $\var\left[E_T\right]$ as obtained with \eqref{avg_pow} and \eqref{var_pow}, respectively, are given in Table \ref{table_parameters_hardware} for completeness. From Figs.~\ref{fig1} and \ref{fig3}, it is seen the numerical results match very well with the theoretical ones for all the investigated scenarios. The small discrepancy observed for $T=2$ vanishes if a smaller step is used in the random walk model.

\begin{table*}[t]
\renewcommand{\arraystretch}{1.}
\caption{Numerical values of $\mathbb{E}\left[E_T\right]$ and $\var\left[E_T\right]$ for the parameter setting in Figs.~\ref{fig1} and \ref{fig3}}
\label{table_parameters_hardware} \vskip2mm
\centering
\begin{tabular}{|c||c|c||c||c|c|}
\hline
 {\bf{Fig.~\ref{fig1} }} &  $\mathbb{E}\left[E_T\right]$ &  $\var\left[E_T\right]$ &   {\bf{Fig.~\ref{fig3} }}  &  $\mathbb{E}\left[E_T\right]$ &  $\var\left[E_T\right]$\\
\hline

$T=2, N=32$ & $1.33$  & $0.015$ & $T=2, \beta = 4$ & $0.889$  & $0.0067$ \\
$T=2, N=64$ & $0.889$  & $0.0067$ & $T=2, \beta = 6$ & $1$  & $0.0029$\\
$T=10, N=32$ & $6.668$ & $0.0891$ & $T=10, \beta = 4$ & $4.445$ & $0.0396$ \\
$T=10, N=64$ & $4.445$ & $0.0396$ & $T=10, \beta = 6$ & $5$ & $0.0169$\\

\hline
\end{tabular}
\end{table*}

%\begin{table}[t]
%\renewcommand{\arraystretch}{1.}
%\caption{Values of $\mathbb{E}\left[E_T\right]$ and  $\var\left[E_T\right]$}
%\label{table_parameters_hardware} \vskip5mm
%\centering
%\begin{tabular}{|c||c|c|}
%\hline
%$K=64, N=128$ &  $\mathbb{E}\left[E_T\right]$ &  $\var\left[E_T\right]$\\
%\hline
%
%$T=2, \beta = 4$ & $1.33$  & $0.015$ \\
%$T=2, \beta = 6$ & $0.889$  & $0.0067$ \\
%$T=10, \beta = 4$ & $6.668$ & $0.0891$\\
%$T=10, \beta = 6$ & $4.445$ & $0.0396$\\
%
%\hline
%\end{tabular}
%\end{table}

\subsection{Application: Dimensioning of Cell Batteries}

A possible application of the results of Theorem 1 is as follows. Assume that the energy level $\eta$ of a battery-powered BS has to be designed such that the achievable rate of each UE is $r_k= \log_2\left(1+{\rho}/{\sigma^2}\right)$ and the probability of running out of energy (before replacement or reloading) is smaller than some given threshold $\epsilon$. Mathematically, this amounts to saying that $\text{Pr}\left({E_T} > \eta \right) \le \epsilon$. From the results of Theorem 1, we have that
\begin{align}\label{outage_1}
\text{Pr}\left({E_T} > \eta \right) = Q\left(\frac{\eta - \mathbb{E}\left[E_T\right] }{\sqrt{\var\left[E_T\right] }}\right)
\end{align}
from which it follows that
\begin{align}\label{outage_1}
\eta \ge {\sqrt{\var\left[E_T\right] }} Q^{-1}\left(\epsilon\right) + \mathbb{E}\left[E_T\right].
\end{align}
Consider for example a cell operating over a bandwidth $B =10$ MHz with a radius $R=1$ km and a cut-off parameter $r_0 =  50$ m. Assume $K =32$, $N=64$ and $\beta = 4$. If the noise power is $\sigma^2 =10^{-14}$ W/Hz and $\rho$ is chosen equal to $3\times10^{-14}$ W/Hz, then the achievable data rate of each UE is $r_k = \log(1+\rho/\sigma^2) = 2$ bit/s/Hz. {Set $\ell = 10$ m and $\xi = 1$ minute such that $D = \ell^2/(4\xi) = 25$ m$^2$/minute.} Assume that the replacing (or recharging) time $T$ is $1$ day. In the above circumstances, from \eqref{avg_pow} and \eqref{var_pow} we obtain $\mathbb{E}\left[E_T\right] = 1.73 \times 10^3\,\rm{J}$ and $\var\left[E_T\right]  = 5.65 \times 10^4 \,\rm{J^2}$. Plugging these values into \eqref{outage_1} and setting $\epsilon = 1 \%$ shows that the battery level must satisfy
\begin{align}
\eta \ge 2.28\times 10^3\,\rm{J}.
\end{align}
Notice that the condition that $\epsilon=1\%$ makes the necessary battery level $\eta$ be substantially higher than $\mathbb{E}\left[E_T\right]$.
It is also worth observing that the above value accounts only for the energy required to transmit the signal $\mathbf{s}(t)$ within the time interval $T$. The design of the battery level must also take into account the energy required for digital signal processing, channel coding and decoding, channel estimation and pre-coding, and so forth (see \cite{Emil13} for more details). However, all these quantities can be somehow quantified off-line and easily added to $\eta$ for a correct design.

\section{Conclusions}

In this work, we have studied the energy consumption dynamics in the large limit for a MIMO cellular network in which the UEs move around according to a Brownian motion model. In particular, we have shown that the energy consumption converges in distribution to a Gaussian random variable and we have computed its mean and variance analytically. We have shown that user mobility plays a key role in determining the fluctuations of energy consumption. Numerical results have been used to show that the analytical expressions yield accurate approximations for different settings. As an application of these results, we have dimensioned a battery-powered BS so as to satisfy a certain probability of running out of energy. The results of this work could also be used to get some insights on how designing the number of cells required to cover a given area while taking into account the implementation costs.

%They could also be extended to compute the energy level consumption considering different traffic loads.

It is worth observing that the simplicity of the analytical results comes from the adoption of a ZF precoding technique at the BS. The use of different techniques (such as the regularized ZF) makes the computations much more involved and is currently under investigation. The way different user mobility models impact energy consumption dynamics is also an ongoing research activity.

\bibliographystyle{IEEEtran}
\bibliography{IEEEabrv,refs}%,C:/Users/ARISLM/ALMDocuments/Dropbox/Work/CurrentWork/bibliography/wireless}

\end{document}